\documentclass[12pt]{article}

\usepackage{graphicx}
\usepackage{epstopdf}
\usepackage{amssymb}

\usepackage{sw20aip}

\input{tcilatex}
\begin{document}

\date{November 21, 2006}
\title{Comment on ``Development and assessment of research-based tutorials on heat
engines and the second law of thermodynamics,'' by Matthew J. Cochran and
Paula R. L. Heron [Am. J. Phys. 74, 734---741 (2006)]\medskip }
\author{Manfred Bucher \\
Physics Department, California State\\
University Fresno\\
Fresno CA 93740-8031\bigskip }
\maketitle

Cochran and Heron\cite{1} test student understanding of heat engines and
refrigerators in terms of the pipeline diagram of heat flows and work,
familiar from textbooks. This diagram, drawing on analogies with fluid flow
between reservoirs, nicely illustrates conservation of energy but is silent
on the other constraint involved: the constancy of entropy, $\Delta S=0$.
Not too surprisingly, about two thirds of the students failed to give
correct responses and explanations. I suggest that this outcome results in
large part from the pipeline diagram's exclusive emphasis on energy
conservation which tends to be misleading.

A generation ago a wedge diagram was proposed in these pages\cite{2} that
simultaneously represents conservation of energy and constancy of entropy in
a Carnot cycle. The wedge diagram also inspired extensions to other cycles,
irreversibility, and the second law of thermodynamics,\cite{3} but it has
not yet been adopted in a physics text. As implied by Cochran and Heron's
study, another whole generation of physics students have since been
instructed with strong visual cues on energy conservation in thermodynamic
cycles but without sufficient caveats about the entropy constraint.

\begin{figure}[htbp] 
   \centering
   \includegraphics[width=2in]{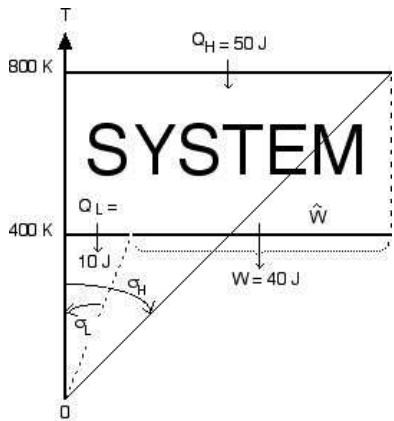} 
   \caption{Wedge diagram of a proposed heat engine}
   \label{fig:example}
\end{figure}

Figure 1 rephrases the same problem of a proposed heat engine that Cochran
and Heron had posed to their students, but here in terms of the wedge
diagram. The amount of heat flows, $Q_{H}$ and $Q_{L}$, and work $W$ are
proportionally represented by the lengths of horizontal lines. The direction
of the vertical arrows relates those quantities to the system (``in'' or
``out'' for $\pm Q$ and $\mp W$), in agreement with standard sign
convention. Conservation of energy, $Q_{H}$ $=-Q_{L}$ $+$ $W$, is shown by
the same top and bottom width of the ``system box'' in the diagram. Entropy
changes are illustrated by angles: The larger angle between the temperature
axis and the solid slanted line, $\tan \sigma _{H}$ $\varpropto $ $\Delta
S_{H}$ $=$ $Q_{H}/T_{H}$, shows that the system's entropy increases more
from the heat inflow $Q_{H}$ than its entropy decrease from the heat
outflow, shown by the smaller angle between the dotted slanted line and the
temperature axis, $\tan \sigma _{L}$ $\varpropto $ $\Delta S_{L}$ $=$ $%
Q_{L}/T_{L}$. This result is in contradiction with the requirement $\Delta
S=0$ for a complete cycle and thus renders the proposed engine impossible.
The maximum work that can be obtained by a \textit{reversible} Carnot engine
is represented in Fig. 1 by the horizontal segment to the right of the solid
slanted line, $\hat{W}$.

\begin{figure}[htbp] 
   \centering
   \includegraphics[width=2in]{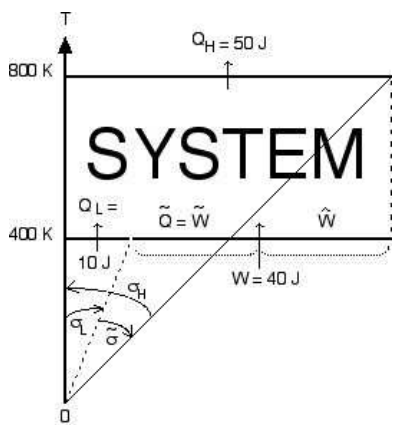} 
   \caption{Wedge diagram of a proposed refrigerator}
   \label{fig:example}
\end{figure}

Cochran and Heron's second examination question reverses the heat flows and
work for a proposed refrigerator. Such a refrigerator proves possible
although of low performance, $\kappa =Q_{L}/W=10J/40J=25\%$. The wedge
diagram in Fig. 2 shows that a certain part of the work $W$---denoted as $%
\tilde{W}$ and represented by the horizontal segment between the dotted and
solid slanted lines---is internally converted to heat, $\tilde{W}
\longrightarrow $\ $\tilde{Q}$, due to irreversible processes. This gives
rise to a combined entropy increase by $\Delta S_{L}+\Delta \tilde{S}$ $%
=Q_{L}/T_{L}+\tilde{Q}/T_{L}$, that is offset by the entropy decrease $%
\Delta S_{H}=Q_{H}/T_{H}$.

If instead the refrigerator were to operate reversibly, then that amount of
heat, $\tilde{Q}$, would be additionally extracted from the low-temperature
reservoir by the work $\hat{W}=W-\tilde{W}$, represented by the horizontal
segment ranging from the solid slanted line to the far right. The
performance of a reversible refrigerator under these conditions would be $%
\hat{\kappa}=(Q_{L}+\tilde{Q})/\hat{W}=T_{L}/(T_{H}-T_{L})=100\%$.

The persistent, almost universal use of the pipeline diagram---a half-truth,
at best---is astounding. To remedy the consequences of its pitfall Cochran
and Heron have developed tutorials\cite{1} that augment this graphical
device by calculations (cycle efficiency or performance) and statements of
the second law of thermodynamics in its various formulations. However, such
a graphic-\textit{cum}-calculational/verbal procedure appears rather
asymmetric. The wedge diagram, on the other hand, implements both conditions
of thermodynamic cycles, $\Delta E=0$ and $\Delta S=0$, graphically on a
equal\ footing.

How can students benefit from the virtues of the wedge diagram while being
exposed in their texts to the pipeline diagram? Several, mutually enhancing
possibilities come to mind: (1) An ``improved,'' wedge-diagram treatment of
the Carnot cycle in lecture. (2) A hand-out sheet with diagram rules and
simple examples. (3) Incorporation of the wedge diagram in tutorials in the
spirit of Cochran and Heron. (4) Use in online teaching and online homework.
(5) Student encouragement (extra credit?) for drawing the diagram---like a
free-body diagram in mechanics---in selected homework problems from the
textbook. (6) Use of the wedge diagram in exams.\bigskip

\noindent \textbf{ACKNOWLEDGMENT}

I thank Preston Jones for help with computer graphics.

\end{document}